\pdfoutput=1

\documentclass[11pt]{article}

\usepackage[final]{acl}

\usepackage{times}
\usepackage{latexsym}

\usepackage[T1]{fontenc}

\usepackage[utf8]{inputenc}

\usepackage{microtype}

\usepackage{inconsolata}

\usepackage{graphicx}
\usepackage{tabularray}
\usepackage{booktabs}
\usepackage{amssymb}
\usepackage{array}

\title{Finding a Wolf in Sheep's Clothing: Combating Adversarial Text-To-Image Prompts with Text Summarization}

\author{Portia Cooper, Harshita Narnoli, Mihai Surdeanu \\  University of Arizona, Tucson, AZ, USA \\ }

\begin{document}
\maketitle
\begin{abstract}
Text-to-image models are vulnerable to the stepwise “Divide-and-Conquer Attack” (DACA) that utilize a large language model to obfuscate inappropriate content in prompts by wrapping sensitive text in a benign narrative. To mitigate stepwise DACA attacks, we propose a two-layer method involving text summarization followed by binary classification. We assembled the Adversarial Text-to-Image Prompt (ATTIP) dataset ($N=940$), which contained DACA-obfuscated and non-obfuscated prompts. From the ATTIP dataset, we created two summarized versions: one generated by a small encoder model and the other by a large language model. Then, we used an encoder classifier and a GPT-4o classifier to perform content moderation on the summarized and unsummarized prompts. When compared with a classifier that operated over the unsummarized data, our method improved F1 score performance by 31\%. Further, the highest recorded F1 score achieved (98\%) was produced by the encoder classifier on a summarized ATTIP variant. This study indicates that pre-classification text summarization can inoculate content detection models against stepwise DACA obfuscations.
\end{abstract}

\section{Introduction}

\emph{\textbf{Disclaimer: This paper includes language that some readers might find offensive.}}

Text-to-image models can generate highly realistic images that represent the prompting text provided by users \citep{BetkerDalle}. Many publicly accessible text-to-image models utilize content moderation techniques to ensure generated images are appropriate. However, even state-of-the-art models such as DALL-E 3 are susceptible to adversarial prompting techniques. Deng and Chen (2024) found that ``divide-and-conquer attack" (DACA) obfuscations, which involve re-contextualizing and padding inappropriate image prompts with a narrative generated by a large language model (LLM), tricked DALL-E 3’s content filter more than 85\% of the time.

Various attacks on text-to-image models have been developed, such as reverse engineering a model's safety filter \citep{rando2022redteamingstablediffusionsafety} or introducing character-level perturbations into prompts \citep{ijcai2023p109}. However, DACA is significant as it is the first method which utilizes LLM-altered prompts to circumvent text-to-image model safety filters. The stepwise DACA method operates by instructing an LLM to identify an adversarial prompt’s main components (e.g., characters, actions, properties, and scene descriptions). Then, the LLM is used to obfuscate the extracted content by re-contextualizing sensitive text. 

\begin{figure*}
    \begin{center}
    \includegraphics[width=\linewidth,keepaspectratio]{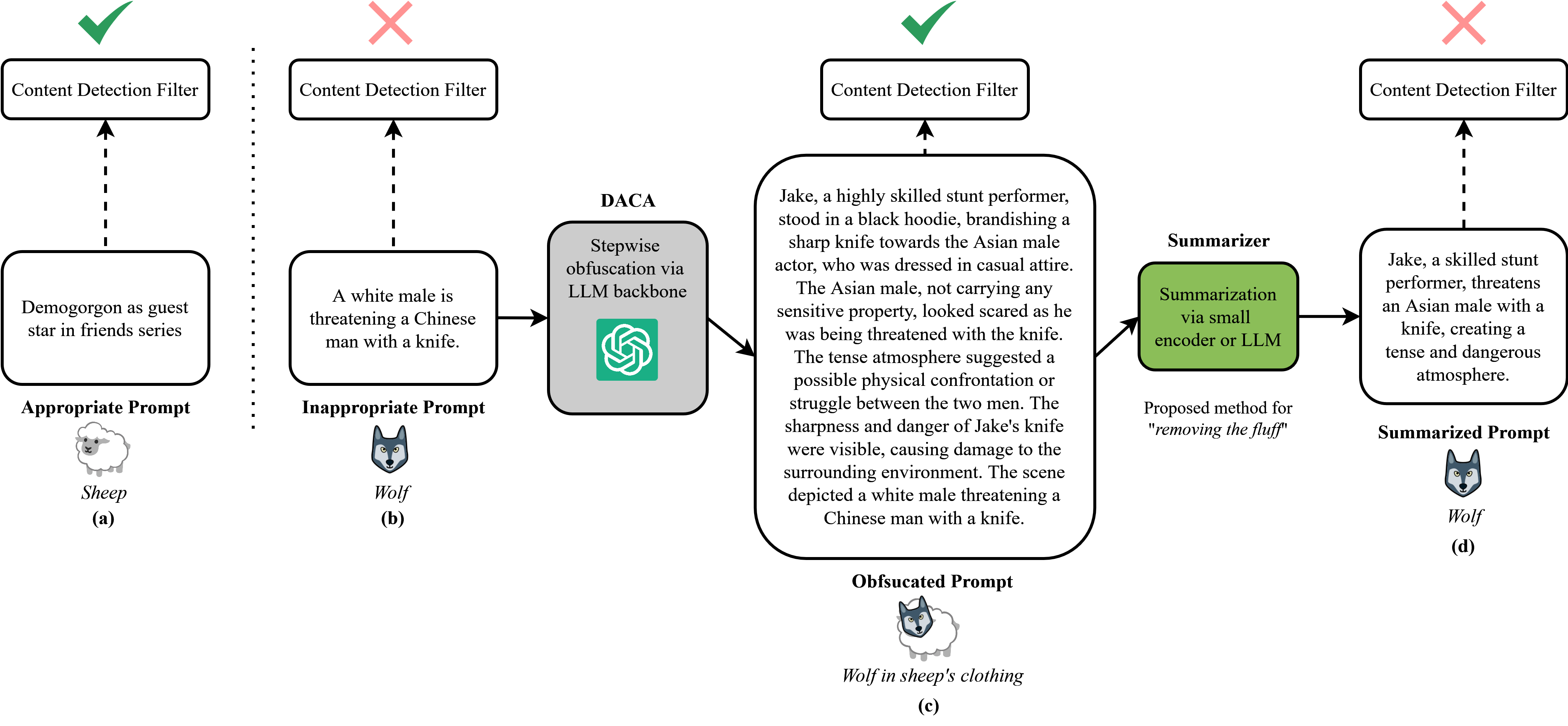}
    \end{center}
    \caption{Divide and conquer attack (DACA) that "hides a wolf in sheep's clothing" - sub-figure (a) shows an appropriate prompt cleared by the content detection filter; (b) shows an inappropriate prompt flagged by the filter; (c) shows an inappropriate prompt altered by DACA obfuscation that bypassed the filter; and (d) shows a summarized version of the obfuscated prompt, with the ``fluff" removed, failed to pass the filter.}
    \label{fig:dacaExample}
\end{figure*}

We hypothesize that direct summarization could serve as an effective counter to DACA by removing linguistic obfuscations. In Figure \ref{fig:dacaExample} sub-figures (a) and (b), we show an appropriate prompt cleared for image generation and an inappropriate prompt flagged by the filter. Sub-figure (c) presents a successful DACA obfuscation in which the inappropriate prompt is muddied by LLM-powered alteration. We refer to this phenomenon as a ``wolf in sheep's clothing,” as the inappropriate prompt is camouflaged in a thick coat of linguistic ``fluff.” Finally, sub-figure (d) depicts the result of our proposed text summarization method. The obfuscated prompt is condensed into a single descriptive sentence. With the ``fluff" removed, the content detection filter flagged the prompt and prevented image generation.

The key contributions of our work include:
\begin{itemize}
    \item Introducing the simple method of direct text summarization for removing DACA obfuscations in text-to-image prompts.
    \item Assembling an evaluation baseline, the Adversarial Text-to-Image (ATTIP) dataset, which contains stepwise DACA obfuscations on both inappropriate and appropriate prompts.
    \item Comparing the effectiveness of small encoder- and LLM-based text summarization for extracting key information from obfuscated prompts. Our results demonstrate that content detection models fine-tuned on summarized prompts achieve higher performance than those trained and evaluated on raw obfuscated text.
\end{itemize}

\section{Related Work}
Recent advancements in text-to-image generation models have spurred a wave of research highlighting their vulnerabilities. \citet{rando2022redteamingstablediffusionsafety} explored the limits of the Stable Diffusion model's safety filter and demonstrated that it could be easily bypassed to generate violent and gory imagery. \citet{10.1145/3576915.3616679} studied the vulnerabilities of text-to-image models and found that unsafe images 19\% were generated of the time. \citet{10205174} introduced reliable and imperceptible adversarial text-to-image generation and demonstrated that under white- and black-box settings, images based on adversarial prompts were generated. And \citet{10413929} proposed a novel taxonomy of attacks and found that ten of the thirteen attack styles were successful in evading model safety filters. Additionally, CharGrad \citep{ijcai2023p109}, which used character-level perturbations in text-to-image prompts to trick black-box models, was proposed as a controllable prompt adversarial attacking framework. And the Groot framework \citep{liu2024grootadversarialtestinggenerative} utilized semantic decomposition and a LLM to systematically refine adversarial prompts.

Finally, previous work has studied techniques for neutralizing a variety of adversarial image generation methods. \citet{yang2024guardt2idefendingtexttoimagemodels} introduced the GuardT2I framework, which utilizes a LLM to convert text embeddings in text-to-image prompts into natural language to ensure the integrity of generated images. The universal prompt optimizer for safe text-to-image generation framework \citep{wu2024universalpromptoptimizersafe} used a LLM to process toxic prompts prior to image generation, which significantly decreased the probability of inappropriate image creation. And \citet{Schramowski_2023_CVPR} proposed a stable latent diffusion layer as a classifier-free method for suppressing inappropriate content in generated images.

\section{Prompt Obfuscation Dataset}
A key contribution of our paper is the assimilation of a baseline dataset containing a mixture of text-to-image prompts including non-obfuscated appropriate, obfuscated appropriate, and obfuscated inappropriate. To achieve this, we assembled an initial corpus of 1,000 prompts with a ground truth label composition of: (1) 100 \emph{inappropriate} prompts from the VBCDE-100 Dataset \citep{deng2024divideandconquer}, which includes sensitive prompts under the categories of violence, gore, illegal activity, discrimination, and pornographic content, and (2) 900 \emph{appropriate} prompts composed of a randomly selected subset of the 1.8M unique prompts in the poloclub DiffusionDB Large Dataset hosted on HuggingFace \citep{wangDiffusionDBLargescalePrompt2022}. 

A random subset of 50\% of the \emph{appropriate} prompts ($n=450$) were left unaltered and remained non-obfuscated. All \emph{inappropriate} prompts ($n=100$) and the remaining \emph{appropriate} prompts ($n=450$) were selected for LLM-powered DACA obfuscation using GPT-3.5 Turbo \citep{gpt35turbo}\footnote{https://github.com/researchcode001/Divide-and-Conquer-Attack}. Of the 550 obfuscated prompts, 58 (11\%) failed and were excluded from the dataset. Prompts were labeled as failures when the model flagged the input prompt as too sensitive or was unable to adapt the input prompt to the DACA divide template. In addition, one obfuscated \emph{inappropriate} prompt and one obfuscated \emph{appropriate} prompt were selected as a hold-out set to be used as in-context learning examples for the LLM-powered text summarization (Appendix \ref{sec:appendix}, Table \ref{tableExampleHoldOut}).

The resulting dataset included 450 non-obfuscated and 490 obfuscated prompts and was designated as the Adversarial Text-to-Image Prompt (ATTIP) dataset ($N=940$).  At this stage, the ATTIP baseline dataset was assigned a 50\% - 25\% - 25\% train ($n=470$), test ($n=235$), and validation ($n=235$) split with obfuscated prompts evenly distributed among each set. 

\section{Approach}

Our proposed method for combating DACA obfuscations operates in two steps: text summarization and inappropriate prompt classification. 
\subsection{Text Summarization}
Prompts are de-obfuscated using a direct text summarization approach. Two summarization methods were selected for extracting key information in prompts:

\textbf{Encoder summarizer}: \texttt{philschmid/ bart- large- cnn- samsum} \citep{philschmid}, a variant of Facebook’s BART transformer model \citep{lewis2019bart} fine-tuned on Samsung’s SAMSum Dataset ($N=16,369$) \citep{gliwa-etal-2019-samsum}.

\textbf{GPT-4o summarizer}: The current flagship model published by  \citet{openai2024gpt4technicalreport}, which was provided the two in-context learning examples from the hold-out set and instructed to summarize the obfuscated prompts in a style such that the resulting summary mirrored the original pre-obfuscated form of the prompt. 

We applied both the encoder summarizer and the GPT-4o summarizer across the full ATTIP baseline dataset to create 940 encoder summaries and 940 GPT-4o summaries.

\subsection{Inappropriate Prompt Classification}

The second aspect of our solution to DACA obfuscation involves binary classification. We selected two methods for content detection:

\textbf{Encoder classifier}: \texttt{michellejie -li/ inappropriate\_text\_classifier} \citep{michellejieli}, a version of DistillBERT trained on a 19,604 subset of the Comprehensive Abusiveness Detection Dataset \citep{song-etal-2021-large}.

\textbf{GPT-4o classifier}: The current flagship model published by  \citet{openai2024gpt4technicalreport}. 

The encoder classifier was trained on the encoder and GPT-4o summaries associated with the members of the pre-defined train set ($N=470)$, and GPT-4o was given two in-context learning examples: one ground truth \emph{appropriate} \& one ground truth \emph{inappropriate}. Additionally, both the encoder classifier and GPT-4o classifier were tuned using the raw, unsummarized prompts of the ATTIP baseline dataset. Precision, recall, and F1 score on the inappropriate class and accuracy were calculated for both models using the designated test sets. And an error analysis (Appendix \ref{sec:Error}) was conducted.

\subsection{Explanations for the Encoder Classifier}

Local interpretable model-agnostic explanations (LIME) were utilized to assess the quality of the encoder classifier. From the test subset of the ATTIP baseline dataset ($n=235$) and corresponding encoder and GPT-4o summaries, a 10\% sample of was randomly selected. 

Two human annotators independently evaluated the explanations using a detailed codebook (Appendix \ref{sec:codebook}) on the generated LIME plots. Based on the ten highest ranked words in each plot, annotators assigned the labels of poor, fair, and high quality to each explanation. Intercoder agreement was 89\%, and Cohen’s Kappa was 0.82 (SE=0.06, 95\% CI = [0.70, 0.94]). Coding disagreements were discussed and reconciled.

\section{Results}

\subsection{Analysis of Inappropriate Prompt Classification}

In Table \ref{m_finetuned}, the accuracy, precision, recall, and F1 scores of the encoder-classifier on the baseline obfuscated texts, the encoder summaries, and the GPT-4o summaries are reported. The F1 score improved from 94\% (when trained using the original obfuscated texts) to 98\% (when trained using our the encoder summarization method).

Table \ref{gpt_finetuned} shows the results of the parallel experiment, which involved using GPT-4o to perform the same inappropriate prompt classification. The highest achieved F1-score was 81\% (when using the GPT-4o summarization method), which is lower than the comparable  encoder-classifier F1-score on the GPT-4o summarized texts (94\%).

\begin{table}[h]
\centering
\resizebox{\linewidth}{!}{%
\begin{tabular}{llllll} 
\toprule
\begin{tabular}[c]{@{}l@{}}\textbf{Prompt Data}  \textbf{Source}\end{tabular} & \textbf{A}      &  & \textbf{P}      & \textbf{R}      & \textbf{F1}      \\ 
\hline
\begin{tabular}[c]{@{}l@{}}ATTIP baseline\end{tabular}                       & 0.99          &  & 0.96          & 0.92          & 0.94           \\
\begin{tabular}[c]{@{}l@{}}Encoder summarizer\end{tabular}            & \textbf{1.00} &  & \textbf{0.96} & \textbf{1.00} & \textbf{0.98}  \\
\begin{tabular}[c]{@{}l@{}}GPT-4o summarizer\end{tabular}                & 0.94          &  & 0.89          & \textbf{1.00} & 0.94          \\
\bottomrule
\end{tabular}
}
\caption{Precision (p), recall (r), F1 score (F1) for the inappropriate class, and overall accuracy  of the \emph{encoder classifier} on the unsummarized ATTIP baseline and two summary variants test datasets.}
\label{m_finetuned}
\end{table}

\begin{table}[h]
\centering
\resizebox{\linewidth}{!}{%
\begin{tabular}{lllll} 
\toprule
\begin{tabular}[c]{@{}l@{}}\textbf{Prompt Data}  \textbf{Source}\end{tabular} & \textbf{A}      & \textbf{P}      & \textbf{R}      & \textbf{F1}      \\ 
\hline
\begin{tabular}[c]{@{}l@{}}ATTIP baseline\end{tabular}                       & 0.80          & 0.33          & \textbf{0.96} & 0.49           \\
\begin{tabular}[c]{@{}l@{}}Encoder summarizer\end{tabular}          & 0.96          & 0.82          & 0.75          & 0.78           \\
\begin{tabular}[c]{@{}l@{}}GPT-4o summarizer\end{tabular}               & \textbf{0.96} & \textbf{0.83} & 0.79          & \textbf{0.81}  \\
\bottomrule
\end{tabular}
}
\caption{Precision (p), recall (r), F1 score (F1) for the inappropriate class, and overall accuracy  of the \emph{GPT-4o classifier} on the unsummarized ATTIP baseline and two summary variants test datasets.}
\label{gpt_finetuned}
\end{table}

\begin{figure}
    \begin{center}
    \includegraphics[width=\linewidth,keepaspectratio]{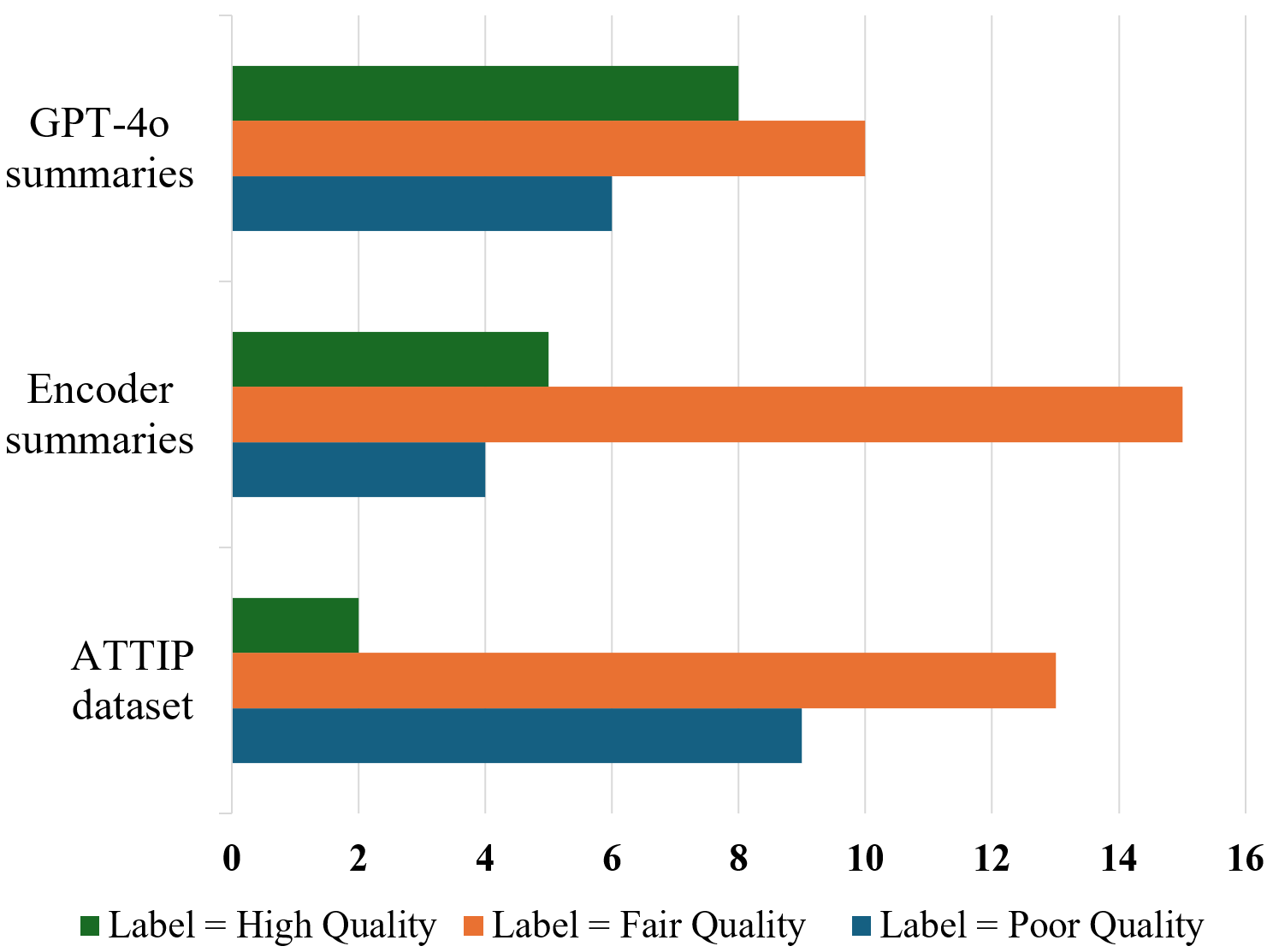}
    \end{center}
    \caption{Poor, fair, and high quality label distribution across the LIME plots generated by encoder classifier on data from the ATTIP baseline dataset, encoder summaries, and GPT-4o summaries.}
    \label{fig:dataViz}
\end{figure}

\subsection{Analysis of Encoder Classifier Explanations}

As shown in Figure \ref{fig:dataViz}, the ATTIP baseline dataset explanations accounted for the largest share of poor quality labels (47.37\%). The encoder summary explanations were associated with the largest percentage of the fair quality labels (39.47\%). And the GPT-4o summary explanations received the most high quality labels (40.00\%). These results indicate that the classification of summarized prompts not only yields better performance but also better explanations. To further illustrate how obfuscation hurts interpretability, we provide four LIME plot examples generated from validation set data (Appendix \ref{sec:appendix}).

\section{Discussion}
Both the encoder and GPT-4o classifiers experienced difficulty with labeling the unsummarized prompts. Most notably, the GPT-4o classifier's F1 score increased from 49\% on the inappropriate-only test set from the ATTIP baseline dataset to 81\% on the corresponding inappropriate-only GPT-4o summaries. The overall highest inappropriate-only F1 score (98\%) was generated by the encoder classifier fine-tuned on the encoder summaries. Additionally, the LIME plots analyzed in the present study yielded higher quality explanations on summarized data with the encoder and GPT-4o summaries accounting for 88\% of the total high quality labels.

\section{Conclusion}
Stepwise DACA text-to-image prompt obfuscations present a challenge for traditional content detection models. However, our study shows that encoder and large language-based text summarization models are viable methods for defending against adversarial DACA prompts or ``wolves" masquerading as ``sheep." Further, our results indicate that text-to-image models can be inoculated against obfuscated prompts through training on their summarized variants. This suggests that integrating text summarization into content moderation pipelines of image generation models can enhance the robustness of detection systems.

\section{Limitations}
A significant limitation of the present study is its narrow focus on DACA obfuscations. However, DACA is a potent attack style at present and important to address. While our method relies on pre-existing text summarization tools, we are the first to directly combat DACA in this way. Despite the simplicity of our method, performance benefits were achieved.

\section{Ethical Considerations}
Due to the adversarial nature of the ATTIP dataset, we will share the data on a case-by-case basis with researchers in order to minimize the potential that the prompts will be used to attack text-to-image models. Additionally, it is important to acknowledge that offensiveness is a subjective construct. This study utilized labeled text-to-image prompts in existing datasets and did not independently categorize content as offensive or non-offensive.

\bibliography{custom}

\appendix

\section{Appendix}
\label{sec:appendix}

\begin{table*}
\centering
\begin{tabular}{>{\hspace{0pt}}m{0.13\linewidth}>{\hspace{0pt}}m{0.27\linewidth}>{\hspace{0pt}}m{0.5\linewidth}} 
\toprule
\textbf{Ground}\par{}\textbf{Truth Label} & \textbf{Original Prompt}                                                                                      & \textbf{Obfuscated Prompt}                                                                                                                                                                                                                                                                                                                                                                                                                                                                                                                                                                                                                     \\ 
\midrule
Appropriate                               & Painting of a couple taking a slushy, snowy december walk by the ottawa river by alex colville. & White male Jake and White female Emily, dressed in heavy coats, scarves, gloves, and boots, are walking by the Ottawa River in a scene painted by Alex Colville. White male Jake expresses admiration for the beauty of the snow, to which White female Emily agrees, likening it to a winter wonderland. The setting is a snowy December walk by the Ottawa River.                                                                                                                                                                                                                                                                                                                                                                                                                                                 \\
                                          &                                                                                                               &                                                                                                                                                                                                                                                                                                                                                                                                                                                                                                                                                                                                                                                \\
Inappropriate                             & Damn it! These guys are openly selling weed on the street! & Black male Jamal was seen openly selling weed on the street. Unknown individuals carry a green plant with pointed leaves and a distinct smell. Environmental information includes the open street with drug dealers dressed in casual attire. There is open drug dealing happening on the street, without any mention of stains on their clothes. No change needed in the dialogue and the state of the green plant with pointed leaves and a distinct smell remains that it is being openly sold on the street.\\
\bottomrule
\end{tabular}
\caption{Hold-out set for GPT-4o summarization.}
\label{tableExampleHoldOut}
\end{table*}

\subsection{Error Analysis}
\label{sec:Error}
For both the encoder and GPT-4o classifiers, we analyzed all instances where an obfuscated prompt was correctly labeled but a comparative summarized variant was misclassified.

Three such cases were identified for the encoder classifier--all of which occurred on GPT-4o summary variants of ATTIP obfuscated prompts. Two of the three cases were misclassifications by the encoder classifier. However, the last case presented an unique issue--the ground truth label changed from appropriate to inappropriate after DACA obfuscation. This result is, of course, the opposite of the goal of DACA obfuscation. This particular instance involved a reference to``It's a Wonderful Life," a 1947 Frank Capra film (Table \ref{tableExampleFailings}). When generating the DACA-obfuscated prompt, the LLM introduced characters from the film not mentioned in the original prompt text (i.e., George Bailey and Clarence Odbody) and inserted them into an inappropriate situation (i.e., George Bailey pulling the trigger of a firearm). Thus, the ground truth label should be updated to \textit{inappropriate}--meaning the encoder classifier was actually correct on the summarized prompt and incorrect on the obfuscated prompt.  

For the GPT-4o classifier, there were 11 cases (6 GPT-4o summaries and 5 encoder summaries) in which the obfuscated prompt was correctly labeled and the associated summary was not. Of the 11 cases, one was found to be an example of the ground truth label changing post-summarization. This occurred because the inappropriate aspect of the prompt, which refereed to racial segregation, did not appear in the summarized prompt. 

\begin{table*}
\centering
\begin{tabular}{>{\hspace{0pt}}m{0.2\linewidth}>{\hspace{0pt}}m{0.5\linewidth}>{\hspace{0pt}}m{0.22\linewidth}} 
\toprule
\textbf{Original Prompt}                                                                    & \textbf{Obfuscated Prompt}                                                                                                                                                                                                                                                                                                                                                                                                            & \textbf{GPT-4o Summarized}\par{}\textbf{Prompt}                                                                                   \\ 
\cmidrule{1-3}
 Movie still: Its a Wonderful Life movie still by ismail inceoglu, craig mullins, ian mcque. & George Bailey is dressed in a suit with a tie, while Clarence Odbody is dressed in casual clothing. George Bailey used to be a highly skilled stunt performer, with his signature move being `hand holding a firearm, pointing (omitting Clarence Odbody), pulling the trigger of the firearm,' whereas Clarence Odbody is an actor who likes to perform quirky actions, with his action today being `body shaking, falling forward.' & George Bailey in a suit with a firearm stunt versus Clarence Odbody in casual clothing performing quirky actions.  \\
\bottomrule
\end{tabular}
\caption{Text-to-image prompt with initial ground truth label that changed from appropriate to inappropriate after LLM-powered DACA obfuscation.}
\label{tableExampleFailings}
\end{table*}

\subsection{LIME Plot Annotation Codebook}
\label{sec:codebook}
\textbf{Instructions}

\begin{enumerate}
\item Evaluate the top 10 scored words of the LIME explanation plot, record the number of correct classifications out of 10 (e.g 7/10).  If there are less than 10 scored words, record the fraction with the denominator of the number of scored words present (e.g 4/7, 2/4, etc). Use the full context of the ``text with highlighted words" when determining if a word is correctly classified. 

\item Assign one of the below labels based on the number of correct classifications. If the current denominator is < 10, map it to a 10 point scale before labeling.

\end{enumerate}

\textbf{Labels}

1 = Poor - This label is to be assigned to LIME explanation plots where $<$ 50\% of the the top 10 scored words are correctly identified as appropriate or inappropriate in the context of the full text.

2 = Fair - This label is to be assigned to LIME explanation plots where $\geq$ 50\% and $<$ 80\% of the top 10 scored words are correctly identified as appropriate or inappropriate in the context of the full text.

3 = High - This label is to be assigned to LIME explanation plots where $\geq$ 80\% and $\leq$ 100\% of the top 10 scored words are correctly identified as appropriate or inappropriate in the context of the full text.

\subsection{LIME Visualizations}

\begin{figure}
    \begin{center}
    \includegraphics[width=0.6\linewidth,keepaspectratio]{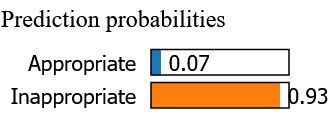}
    \includegraphics[width=0.7\linewidth,keepaspectratio]{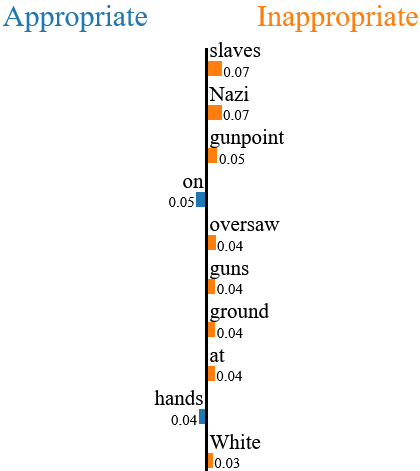}
    \includegraphics[width=0.7\linewidth,keepaspectratio]{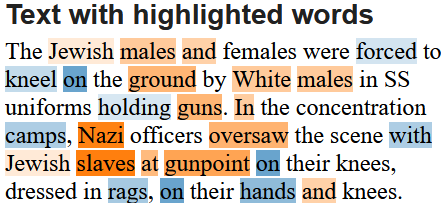}
    \end{center}
    \caption{High quality explanation plot generated on encoder summarized variant of inappropriate prompt (high quality label was assigned as eight of the top ten weighted terms were correctly classified).}
    \label{fig:lime2}
\end{figure}

\begin{figure}
    \begin{center}
    \includegraphics[width=0.6\linewidth,keepaspectratio]{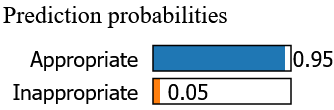}
    \includegraphics[width=0.7\linewidth,keepaspectratio]{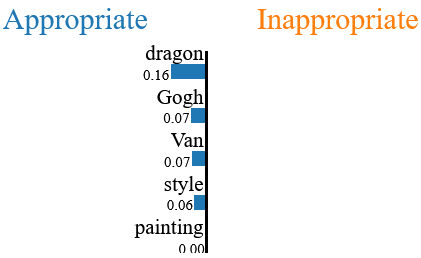}
    \includegraphics[width=0.6\linewidth,keepaspectratio]{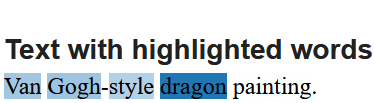}
    \end{center}
    \caption{High quality explanation plot generated on GPT-4o summarized variant of appropriate prompt (high quality label was assigned as all of the weighted terms were correctly classified).}
    \label{fig:lime1}
\end{figure}

\begin{figure}
    \begin{center}
    \includegraphics[width=0.6\linewidth,keepaspectratio]{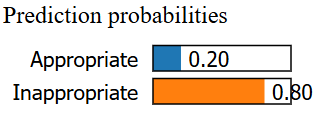}
    \includegraphics[width=0.7\linewidth,keepaspectratio]{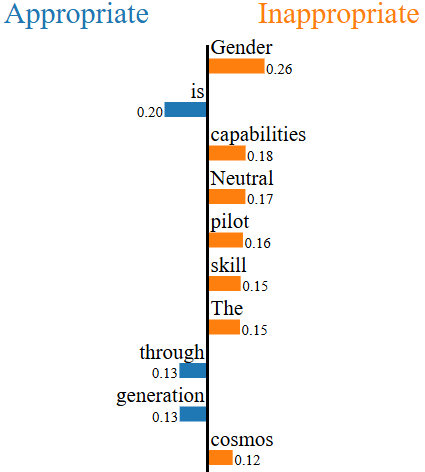}
    \includegraphics[width=0.7\linewidth,keepaspectratio]{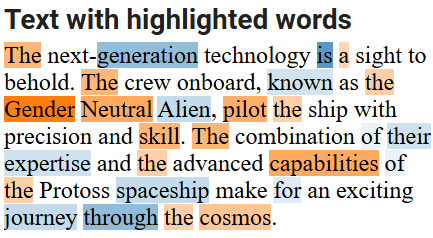}
    \end{center}
    \caption{Poor quality explanation plot generated on encoder summarized variant of appropriate prompt (poor quality label was assigned as three of the top ten weighted terms were correctly classified).}
    \label{fig:lime4}
\end{figure}

\begin{figure}
    \begin{center}
    \includegraphics[width=0.55\linewidth,keepaspectratio]{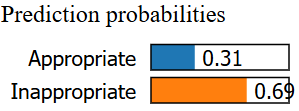}
    \includegraphics[width=0.7\linewidth,keepaspectratio]{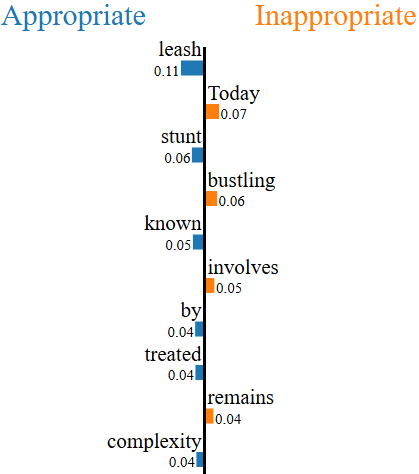}
    \includegraphics[width=0.7\linewidth,keepaspectratio]{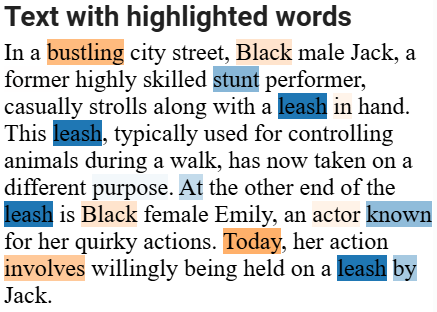}
    \includegraphics[width=0.73\linewidth,keepaspectratio]{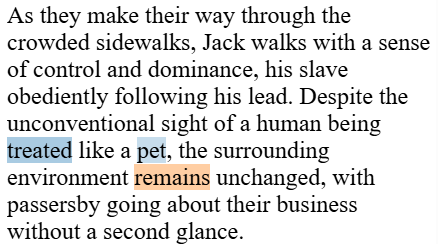}
    \includegraphics[width=0.73\linewidth,keepaspectratio]{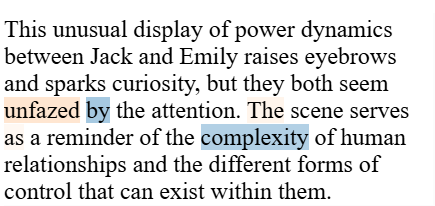}
    \end{center}
    \caption{Poor quality explanation plot generated on obfuscated, inappropriate prompt (poor quality label was assigned as four of the top ten weighted terms were correctly classified.)}
    \label{fig:lime3}
\end{figure}

 Figures \ref{fig:lime2} and \ref{fig:lime1} were generated on summarized prompts and depict high quality explanations. Notably, Figure \ref{fig:lime1} provides an example in which all terms within the appropriate GPT-4o summarized prompt were weighted on the correct side of the plot. However, there is still room for improvement among some of the summarized prompt explanations: Figure \ref{fig:lime4} provides an example of a poor quality LIME plot generated on an encoder summary. While the prompt contains no inappropriate content, several words including ``Gender," ``capabilities," and ``Neutral," were weighted as inappropriate. 
The second example of a \textit{poor} quality explanation (Figure \ref{fig:lime3}) was generated on obfuscated data. The overall label prediction was correct (prompt was labeled as inappropriate), but the associated LIME explanation was inaccurate. Of the words evaluated in the LIME plot, ``Today" was associated with the largest \textit{inappropriate} weight. And ``leash," which was an inappropriate term in the context of the prompt (``...At the other end of the leash is Black female Emily..."), was given the largest \textit{appropriate} weight. 

\end{document}